\documentclass[11pt]{article}
\usepackage{amsfonts,enumerate,amssymb,amsmath}
\usepackage{epsfig}
\usepackage{subfigure,bm,cite}
\usepackage[applemac]{inputenc}
\usepackage{color}

\begin{document}
\thispagestyle{empty}

\title{Multiscale Fractal Descriptors Applied to Nanoscale Images}

\author{
{\bf Jo\~{a}o B. Florindo$^1$}\\ {\bf Mariana S. Sikora$^2$}\\ {\bf Ernesto C. Pereira$^2$}\\ {\bf Odemir M. Bruno$^1$}\\
$^1$ Physics Institute of S\~{a}o Carlos, University of S\~{a}o Paulo\\ S\~{a}o Carlos, SP, Brazil\\
jbflorindo@ursa.ifsc.usp.br, bruno@ifsc.usp.br\\ \\
$^2$ Interdisciplinary Electrochemistry and Ceramics Laboratory\\ Federal University of S\~{a}o Carlos, S\~{a}o Carlos, SP, Brazil\\
ma\_sikora@hotmail.com, decp@power.ufscar.br	
}					 
\date{}					 
\maketitle

\noindent{}{\bf\large Abstract --} 
This work proposes the application of fractal descriptors to the analysis of nanoscale materials under different experimental conditions. We obtain descriptors for images from the sample applying a multiscale transform to the \textcolor{black}{calculation} of fractal dimension of a surface map of such image. Particularly, we have used \textcolor{black}{the} Bouligand-Minkowski fractal dimension. We applied these descriptors to discriminate between two titanium oxide films prepared under different experimental conditions. Results demonstrate the discrimination power of proposed descriptors in such kind of application.
\\

\noindent{}{\bf\large Keywords --}
Porous Titanium Oxide, Nanomaterials, Pattern Recognition, Fractal Dimension, Fractal Descriptors.
\\

\section{Introduction}

\textcolor{black}{Morphological} material characterization is a very important and challenging task throughout Material Science, \textcolor{black}{since} such properties are determinant in the suitability of a given material to a specific application. Among several materials studied recently, a great \textcolor{black}{deal attention} has been given to photoactive ones, such as titanium oxide. Titanium oxide films prepared electrochemically can show a wide variation in their morphology depending on the experimental conditions in which they are prepared. As an example, the morphology can range from self-organized nanoporous 
\cite{Zlamal2007,Ghicov2005,Tsuchiya2005} up to nanostructures without pores definition \cite{Brunella2007,Cai2006}. As \textcolor{black}{described}, many authors \textcolor{black}{have shown} that  morphology is an important aspect to be considered in the photoactive properties of TiO$_{2}$ films \cite{Brunella2007,Macak2005}. \textcolor{black}{Recently} we have shown that the photoactivity of TiO$_{2}$ films \textcolor{black}{prepared} by galvanostatic anodization is strongly affected by the morphology on the early stages of \textcolor{black}{the} anodization process \cite{Sikora2011}. In \cite{Sikora2011} we \textcolor{black}{have} investigated the effect of morphology using an Image Analysis Method, which is based on the computational analysis of visual attributes, such as color, shape and texture. Among these attributes,  texture is a powerful characterizer for material analysis. Although \textcolor{black}{the concept of} texture has no precise definition, it may be comprehended as the spatial organization of pixels in a digital image. A physical consequence of this definition is that this attribute is capable \textcolor{black}{of} express characteristics such as luminosity and roughness of a digitalized object. In this way, it allows a robust description and discrimination of material images, particularly TiO$_{2}$ samples used in this study, which show few or no information about \textcolor{black}{the} pore diameter, but have a rich morphology related to texture information.

Unlike conventional textural data, natural textures \textcolor{black}{do} not present any evident quasi-periodic structure, but persistent random patterns\cite{HDL06}. Among the approaches employed for texture analysis, fractal methods are proposed as the best solution \cite{K99}. These methods, based on fractal dimension \cite{M68} or multifractal spectrum \cite{H01}, can measure the complexity of an object texture, which corresponds to the shape irregularity, also related to the spatial occupation of an object. Therefore, fractal measurements are capable of quantifying the texture homogeneity, allowing a comparison among their information and the consequent discrimination of original materials. Using as example the materials here investigated, some authors \cite{Provata1998,Xagas1999} have shown that the fractal dimension of TiO$_{2}$ films can also be correlated to photoactivity properties of these materials. However, in that papers, the analysis is valid for very regular morphologies, showing deviations when more complex surfaces are considered. 

Although \textcolor{black}{the} fractal dimension is a good descriptor to characterize a texture image, it is inefficient in applications involving the discrimination of a large amount of objects. In fact, it is easy to find objects with different aspects presenting the same fractal dimension \cite{K99}. In order to solve this drawback, the literature presents approaches which extract a lot of descriptors based on fractal geometry, such as Multifractals \cite{H01} and Multiscale Fractal Dimension (MFD) \cite{MCSM02,FCB10,BPFC08}. In the literature, Backes et al. \cite{BCB09} applied MFD in texture analysis employing volumetric Bouligand-Minkowski MFD (VBMFD) to extract a set of descriptors from natural textures with very good results. VBMFD is based on the intensity image mapping onto a 3D surface. In the following, this surface is dilated by a variable radius $r$ and the volume $V(r)$ is calculated for each radius. This expansion process gives a precise measure of the pixel arrangement. As the radius $r$ grows, an interaction among dilation spheres is observed, interfering in $V(r)$ value. Therefore, these values capture changes in spatial distribution of textures along different scales. In this way, VBMFD uses $V(r)$ values as descriptors for a texture, capable of distinguishing different textures with their different spatial arrangement.

Considering the exposed above, in this work it is proposed the use of VBMFD descriptors to discriminate among nanoscale TiO$_{2}$ films prepared electrochemically using two different experimental conditions and characterized by field emission gun scanning electron microscopy (FEGSEM) . In order to demonstrate the efficiency of the proposed method, this work obtains VBMFD descriptors from these films.

\section{Materials and Methods}

\subsection{Fractal Theory}

Since its introduction by Mandelbrot \cite{M68}, the literature presents a lot of works using fractal theory to describe and discriminate several kinds of materials \cite{Provata1998,Xagas1999}. Most of them use fractal dimension as a descriptor of the original samples. This is explained by the fact that fractal dimension measures the complexity of \textcolor{black}{the} object, related to the irregularity or the spatial occupation of that object. This property is strongly correlated to physical important properties.

\subsubsection{Fractal Dimension}

The fractal dimension \cite{M68} is the most commonly used measure to characterize a fractal object. Despite its importance, we cannot find a unique definition for this concept. The most ancient definition corresponds to the Hausdoff-Besicovitch dimension.

If $X\in\Re^{n}$ is a geometrical set of points, the Hausdoff-Besicovitch dimension $dim_{H}(X)$ is given by: 
\begin{equation}
dim_{H}(X)=\inf\left\{ s:H^{s}(X)=0\right\} =\sup\left\{ H^{s}(X)=\infty\right\} ,
\end{equation}
where $H^{s}(X)$ is the $s$-dimensional Hausdorff measure, given by: 
\begin{equation}
H^{s}(X)=\lim_{\delta\rightarrow0}\inf\left\{ \sum_{i=1}^{\infty}{|U_{i}|^{s}:{U_{i}\mbox{ is a \ensuremath{\delta}-cover of X}}}\right\},
\end{equation}
where $| |$ expresses a diameter in $\Re^{n}$, that is, $|U|=sup{|x-y|:x,y\in U}$.

Otherwise, the Hausdorff-Besicovitch definition application could be impracticable in many real situations. This is the case of discrete objects represented in a digital image, as those studied here. For such applications, we have an alternative definition of fractal dimension, which is a generalization from the topological dimension. Thus, the fractal dimension $D$ is provided by: 
\begin{equation}
D(X)=\lim_{\epsilon\rightarrow0}\frac{log(N(\epsilon))}{log(\frac{1}{\epsilon})},
\end{equation}
where $N(\epsilon)$ is the number of objects with linear size $\epsilon$ required to cover the whole object $X$ \cite{F85}. More generically, $N(\epsilon)$ may be considered a measure which varies according to the scale $\epsilon$. Such measure is characterized by a power-law relation with the scale \cite{R94}. A lot of fractal dimension methods were developed by using distinct measures, such as Bouligand-Minkowski \cite{F85}, box-counting \cite{F85}, Fourier \cite{R94}. Here, we are focused on the Bouligand-Minkowski fractal dimension.

\subsubsection{Bouligand-Minkowski}

The original definition of Bouligand-Minkowski fractal dimension $\dim_{B}(X)$ depends on a symmetrical structuring element $Y$: 
\begin{equation}
dim_{M}(X,Y)=\inf\left\{ \tau,meas_{M}(X,Y,\tau)=0\right\} ,
\end{equation}
where $meas_{M}$ is the Bouligand-Minkowski measure: 
\begin{equation}
meas_{M}(X,Y,\tau)=\lim_{r\rightarrow0}\frac{V(\partial X\oplus rY)}{r^{n-\tau}},
\end{equation}
where $r$ is the radius of $Y$ and $V$ is the volume of the dilation between $Y$ and the \textcolor{black}{boundary} $\partial X$ of $X$.

By using neighborhood techniques we may find a simplified version of this dimension, which eliminates the explicit dependence from $Y$:
\begin{equation}
dim_{M}(X)=\lim_{\epsilon\rightarrow0}\left(D_{T}-\frac{logV(X\oplus Y_{\epsilon})}{log\epsilon}\right).\label{eq:BM}
\end{equation}
For the cases such as those studied here, when $X\in\Re^{3}$, the topological dimension is $D_{T}=3$ and $Y_{\epsilon}$ is a sphere with diameter $\epsilon$. For each value of $\epsilon$, each point in $X$ is dilated by $Y_{\epsilon}$ and \textcolor{black}{the} number of points inside the dilated \textcolor{black}{structure} is the dilation volume $V$.

\subsection{Proposed Method}

\begin{figure}[!h]
\centering 
\includegraphics[width=1.0\textwidth]{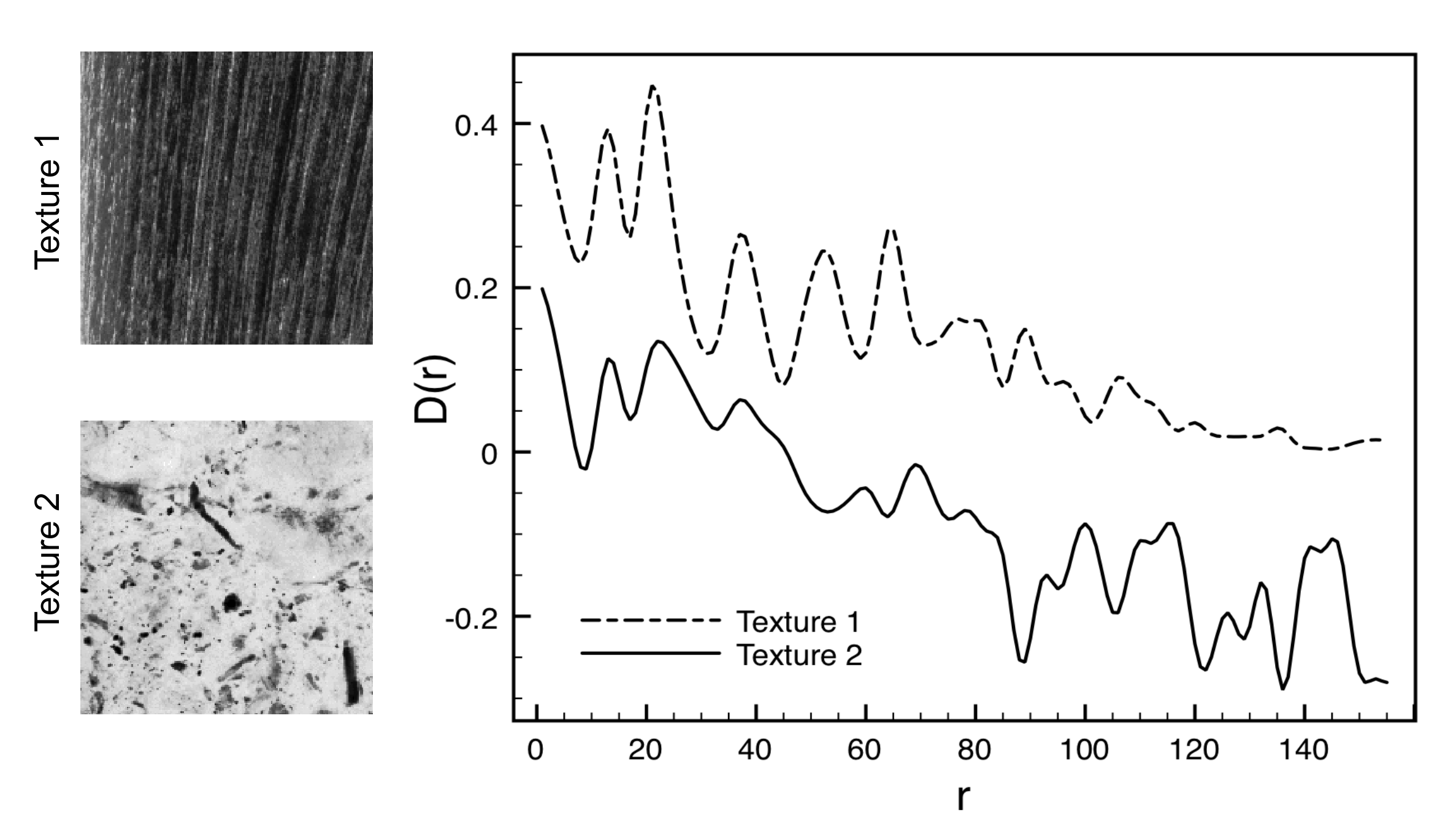} 
\caption{Two textures whose surfaces present the same fractal dimension (FD = 2.014) and their multiscale fractal descriptors $D(r)$, where $r$ is the descriptor index.}
\label{fig:MFD} 
\end{figure}

A significative drawback of fractal dimension is that it is only a unique real number. Therefore, it is possible to find \textcolor{black}{a great} number of objects which despite its same fractal dimension present \textcolor{black}{appearance} completely diverse. This is exemplified in the Figure \ref{fig:MFD} using textures from Brodatz data set \cite{B66}. The authors in \cite{MCSM02} proposed the Multiscale Fractal Dimension technique to solve this issue. In MFD, the fractal dimension is calculated for the object observed under different spatial scales and each value is used as a descriptor for the object. In this work, we propose the analysis of nanoscale FEG images by applying a multiscale approach to the Bouligand-Minkowski fractal dimension.

Initially, we map the image $Img\in[1:M]\times[1:N]\rightarrow\Re$ onto a threedimensional surface 
\begin{equation}
Surf=\{i,j,f(i,j)|(i,j)\in[1:M]\times[1:N]\},
\end{equation}
such that: 
\begin{equation}
f(i,j)=\{1,2,...,max\_gray\}|f=Img(i,j),
\end{equation}
where $max\_gray$ is the maximum pixel intensity.

In the following, the surface is dilated by a variable radius $r$, as depicted in Figure \ref{fig:dilat}. Then the dilation volume $V(r)$ for each dilation radius is calculated. The value of $V(r)$ also corresponds to the number of points with a distance at most $r$ from the object. Therefore, the exact Euclidean distance transform (EDT) \cite{FCTB08} becomes a powerful and efficient tool for this calculus. 

In the threedimensional space, EDT may be defined as the distance of each point in the space to a subset of it. In our case, this subset is the surface and the EDT for each point outside $Surf$ is given by:  
\begin{equation}
EDT(p)=\min\{d(p,q)|q\in Surf^{c}\},
\end{equation}
where $d$ is the Euclidean distance.

In particular, in exact EDT, the distances present discrete values $E$: 
\begin{equation}
E={0,1,\sqrt{2},...,l,...},
\end{equation}
where 
\begin{equation}
l\in D=\{d|d=(i^{2}+j^{2})^{1/2};i,j\in\mathbb{N}\}
\end{equation}

The dilation volume is given through: 
\begin{equation}
V(r)=\sum_{i=1}^{r}{Q(i)},
\end{equation}
where 
\begin{equation}
Q(r)={(x,y,z)|g_{k}(P)-[g_{r}(P)\cap\cup_{i=0}^{r-1}g_{i}(P)]},
\end{equation}
such that: 
\begin{equation}
g_{r}(P)=\left\{ (x,y,z)|[(x-P_{x})^{2}+(y-P_{y})^{2}+(z-P_{z})^{2}\right\} ,
\end{equation}
where 
\begin{equation}
P={(x,y,z)|f(x,y,z)\in Surf}
\end{equation}

The authors in \cite{BCB09} propose the use of values $V(r)$ as descriptors for texture. Such technique is \textcolor{black}{named} Volumetric Bouligand-Minkowski Multiscale Fractal Dimension (VBMFD). In that work, the authors apply this technique to the analysis of plant leaves, achieving interesting results. It is important to stress out that $V(r)$ is directly related to the Bouligand-Minkowski dimension for maximum radius $r$. Besides, each radius corresponds to an observation scale, from further (greater radius) to closer (smaller radius).

This work proposes the application of VBMFD descriptors for the analysis of two TiO$_{2}$ samples prepared under different experimental \textcolor{black}{conditions}. The goal is to discriminate between the samples based on images extracted from different regions of each film.

\subsection{Sample Preparation}

The samples were prepared as follows. Titanium sheet (Alfa Aesar, 99.99\%, 0.25mm thick) with an exposed area of 1 cm$^{2}$, and \textcolor{black}{two} platinum sheets were used as working and counter electrodes, respectively. Before the anodization process, the working electrode was polished with \#1000 SiC and then with \#1200 SiC emery paper followed by vigorous washing with deionized water.

The experiments were carried out under galvanostatic conditions with a home-made current source, measuring the potential difference between working electrode and counter electrode with an HP 34410A multimeter coupled to a computer by an in-house software routine developed with HP-VEE 5.0 software. The two samples investigated were prepared in oxalic acid (sample 1 in a concentration of 0.05 $mol$ $L{}^{-1}$, and sample 2 in 0.5 $mol$ $L{}^{-1}$) applying a constant current density of 10 $mA\, cm^{-2}$ and 20 $mA\, cm^{-2}$ (sample 1 and sample 2, respectively) on the working electrode. The temperature was kept constant at 10 \textdegree{C}  during the anodization of both samples.

After the preparation, the samples were morphlogically characterized using a field emission gun scanning electron microscope (Supra35 - Zeiss). From both samples \textcolor{black}{we have extracted} ten images from different regions of the material in order to collect a representative amount of data. 

\section{Results}

\begin{figure}[!h]
\centering 
\includegraphics[width=1.0\textwidth]{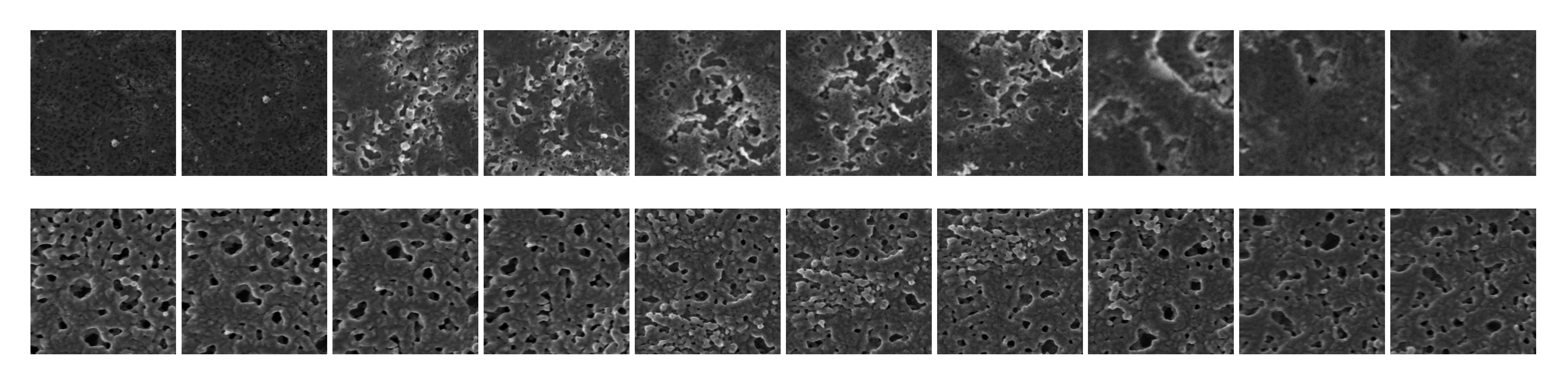}
\caption{Texture images extracted from titanium oxide under two different conditions (above and below).}
\label{fig:img} 
\end{figure}

The fractal descriptors were used \textcolor{black}{for} the discrimination of two TiO$_{2}$ films prepared under different experimental \textcolor{black}{conditions}. From each sample\textcolor{black}{, we have extracted} two micrographs, represented in the Figure \ref{fig:img}.

\begin{figure}[!h] 
\centering 
\mbox{\subfigure{\epsfig{figure=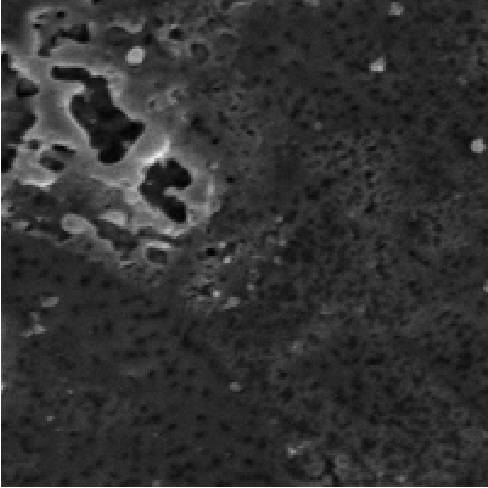,width=3cm, height=3cm}}
      \subfigure{\epsfig{figure=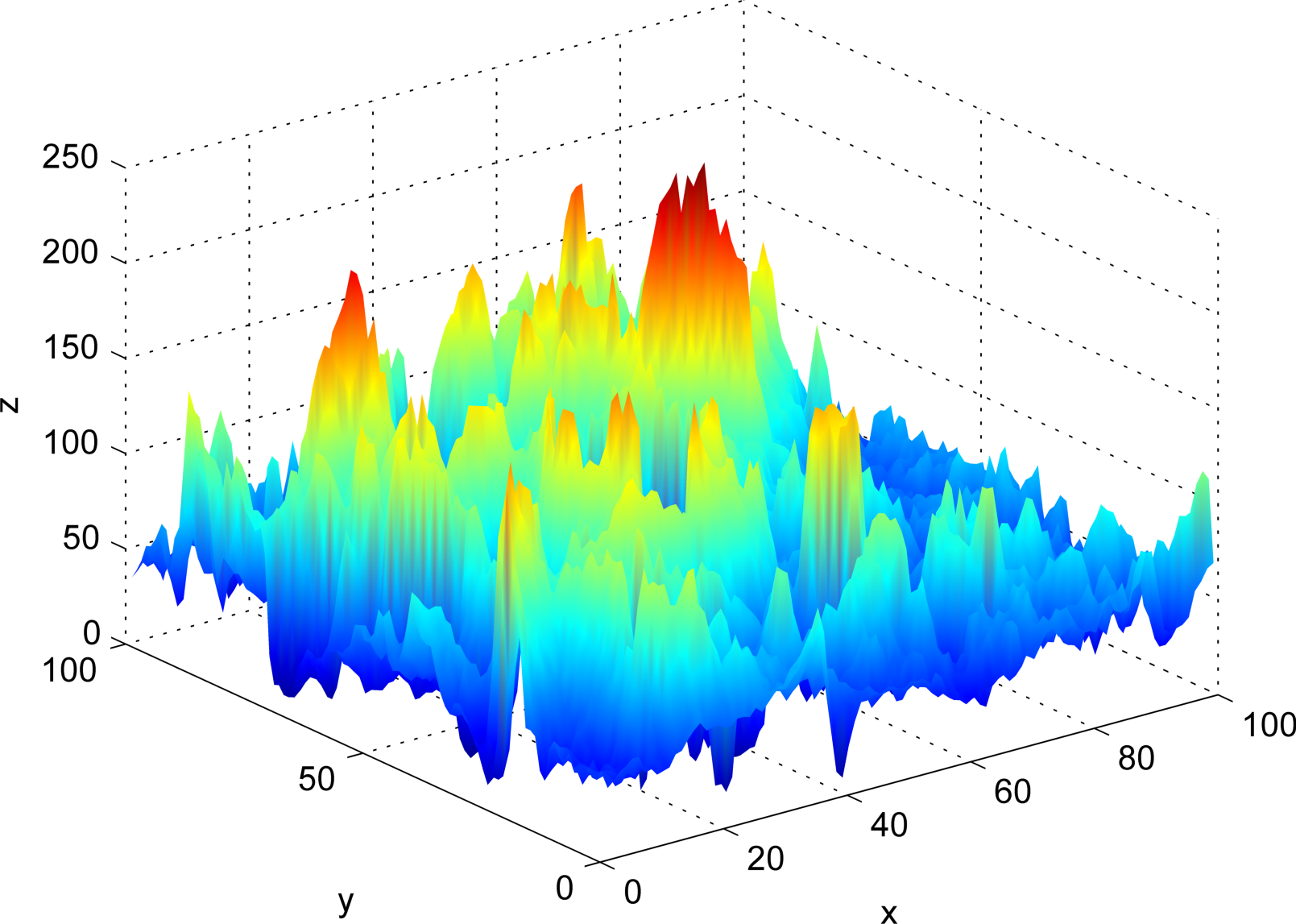,width=6cm, height=4cm}}}
\caption{Texture mapped onto a surface. (a) Original texture. (b) Threedimensional surface.}
\label{fig:surf} 
\end{figure}

Initially, we apply the transform which maps the intensity image onto a surface, as illustrated in the Figure \ref{fig:surf}. 

\begin{figure}[!h]
\centering 
\includegraphics[width=0.4\textwidth]{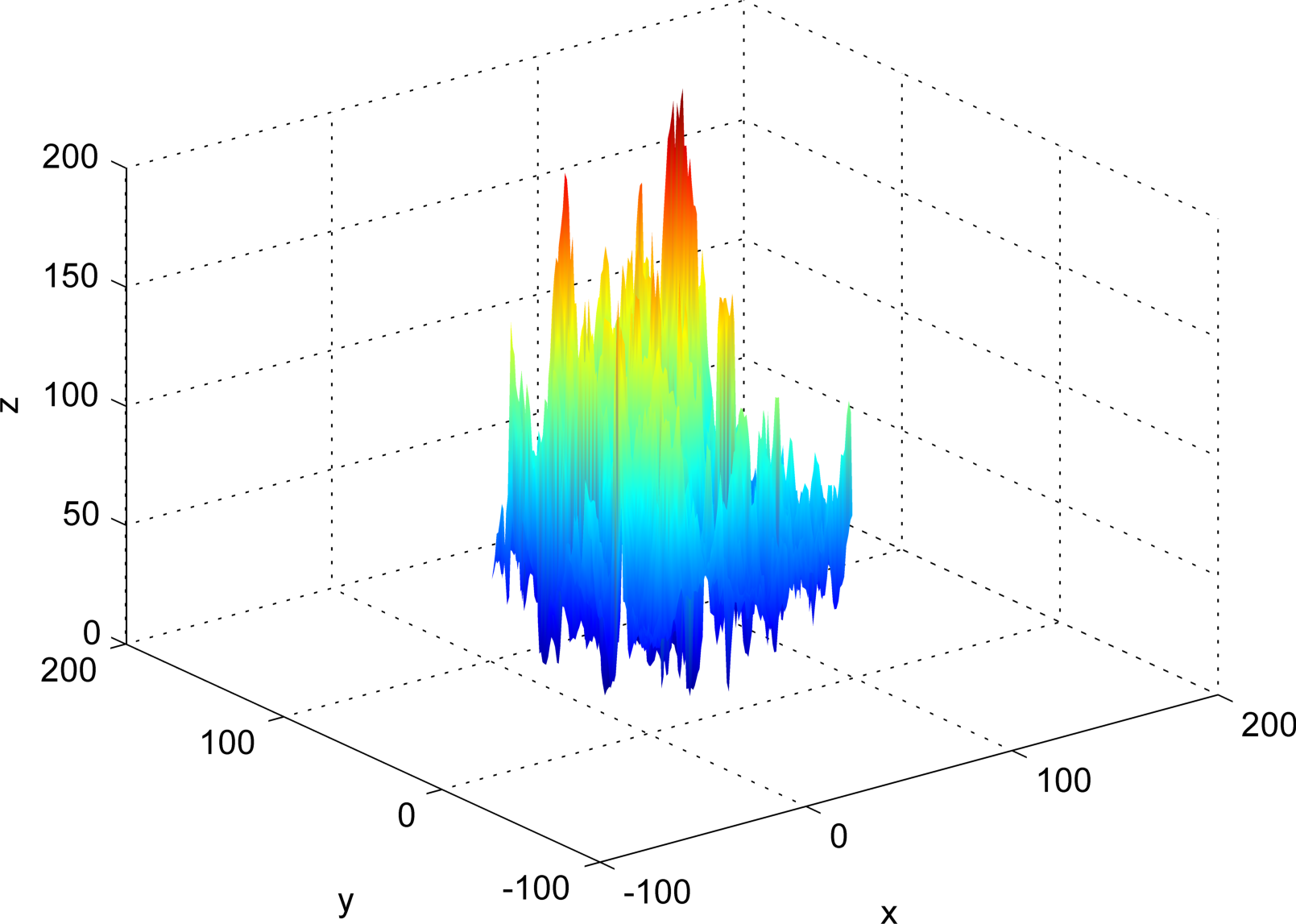}
\includegraphics[width=0.4\textwidth]{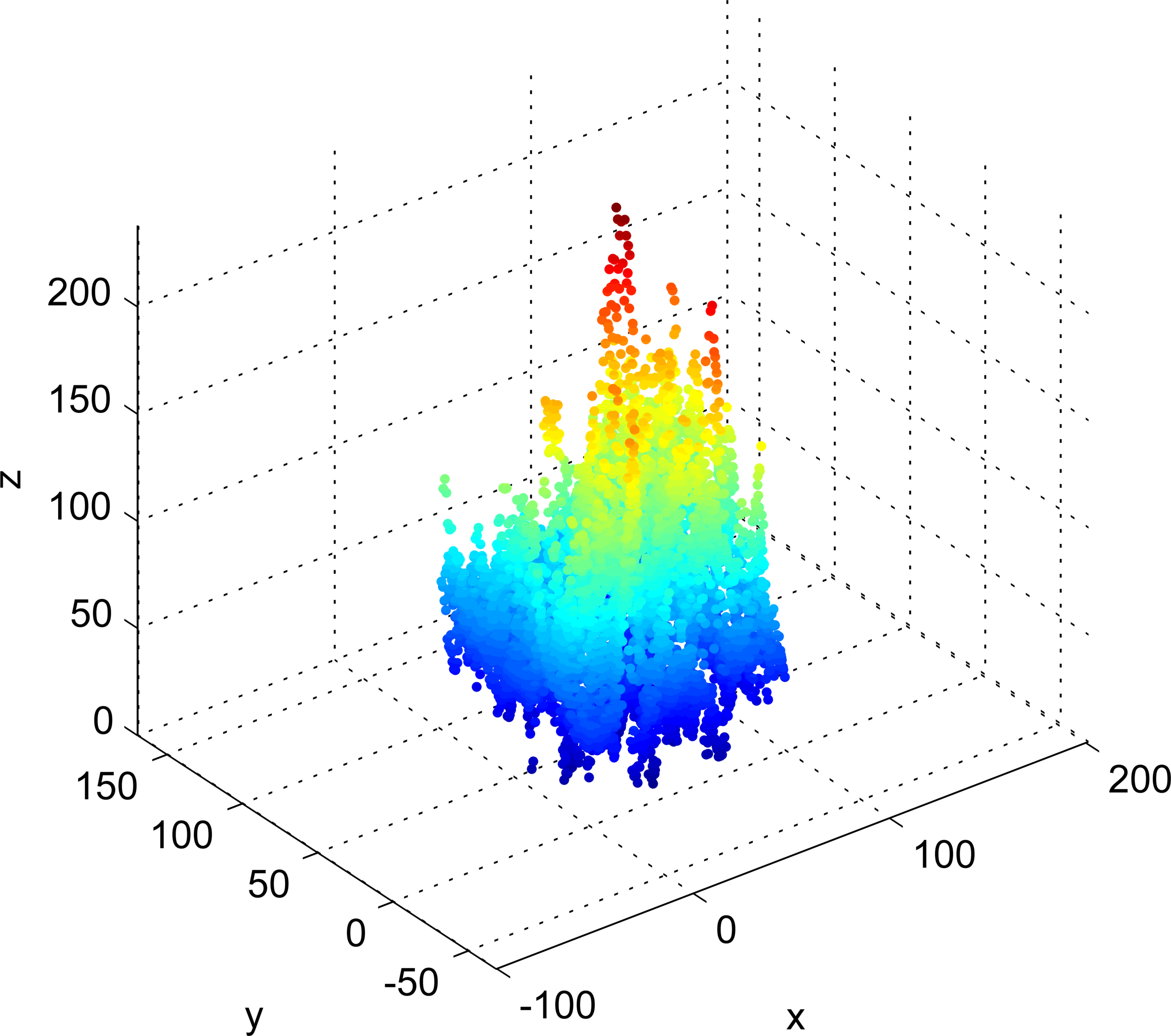} \\
\includegraphics[width=0.4\textwidth]{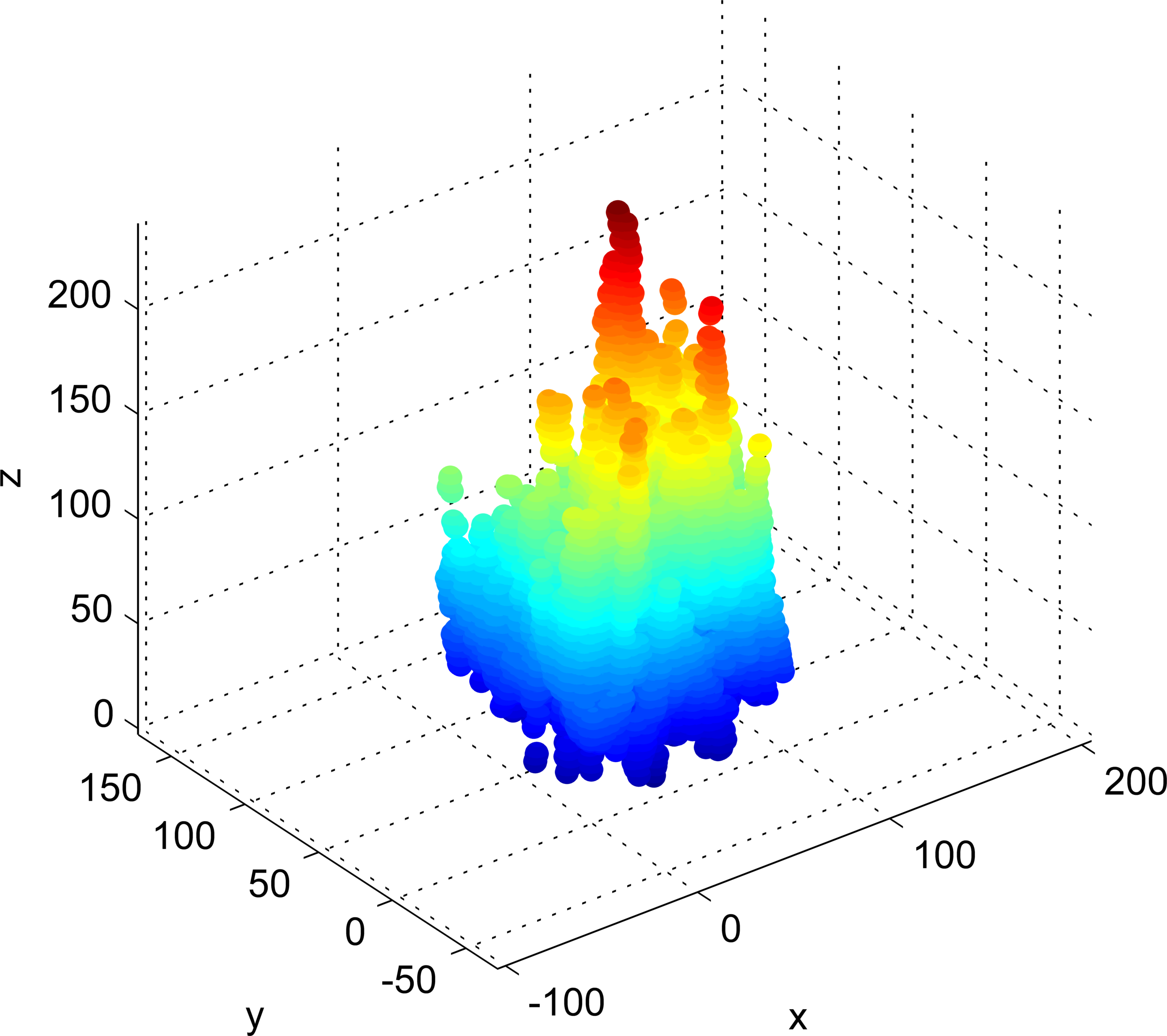}
\includegraphics[width=0.4\textwidth]{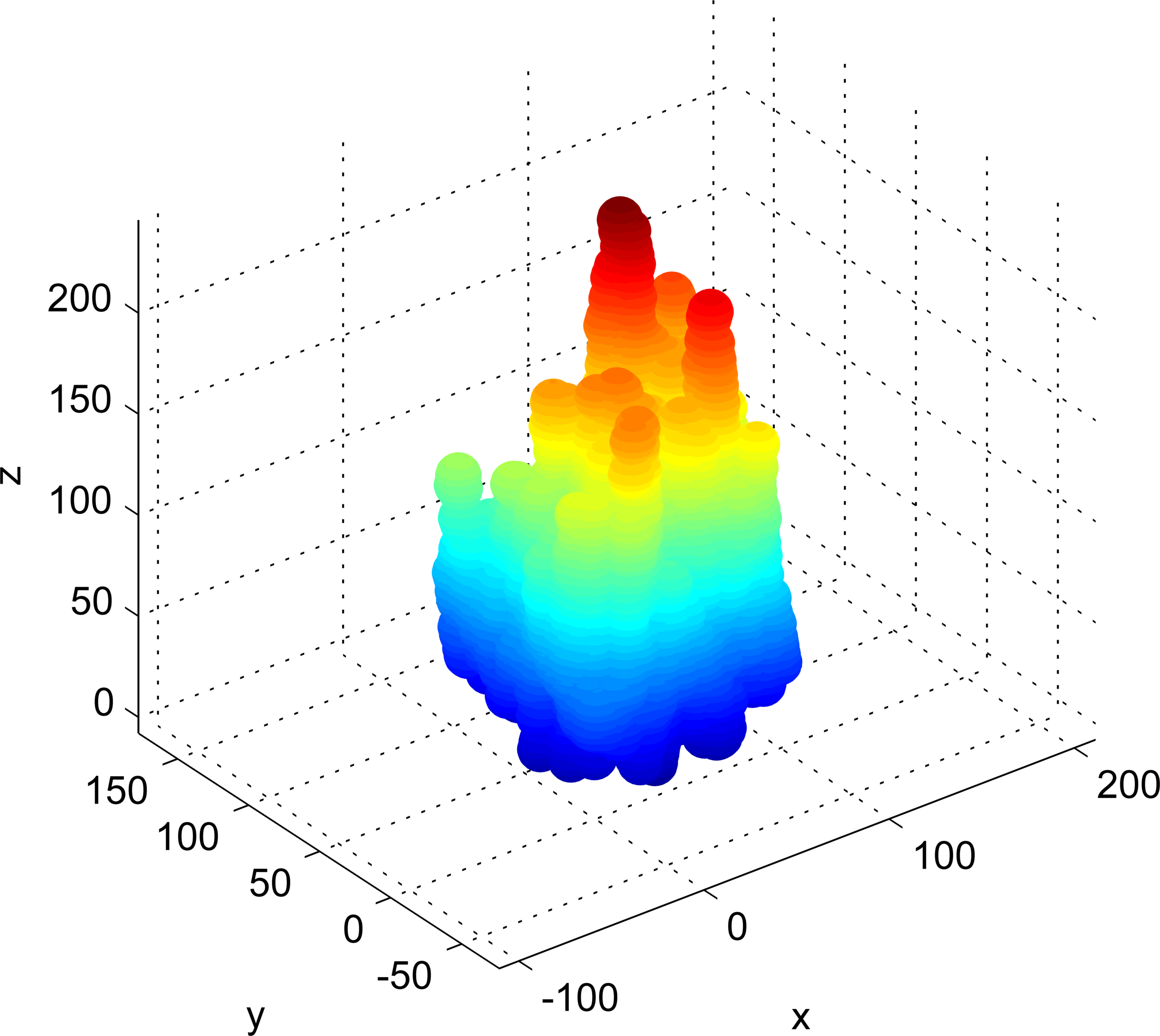}
\caption{Dilated surfaces with different radii. (a) Original surface. (b) Radius 2. (c) Radius 5. (d) Radius 10.}
\label{fig:dilat} 
\end{figure}

Next, the surface is dilated by a variable radius $r$, as depicted in the Figure \ref{fig:dilat}. The conventional Bouligand-Minkowski dimension technique is extracted from the curve $\log(V(r))\times r$, in which, $V(r)$ is the diation volume. Here, the descriptors obtained by the VBMFD technique are provided by:
\begin{equation}
D'(r) = \log(V(r))
\end{equation}

In order to \textcolor{black}{eliminate redundancies} from descriptors $D'_{k}(r)$ of the $k^{th}$ image the following operation was performed: 
\begin{equation}
D_{k}(r)=D'_{k}(r)-\sum_{i=1}^{n}{D'_{i}(r)}/n,
\end{equation}
where $n$ is the total number of images in the dataset and $D_{k}(r)$ is the used descriptor.

\begin{figure}[!h]
\centering 
\includegraphics[width=0.9\textwidth]{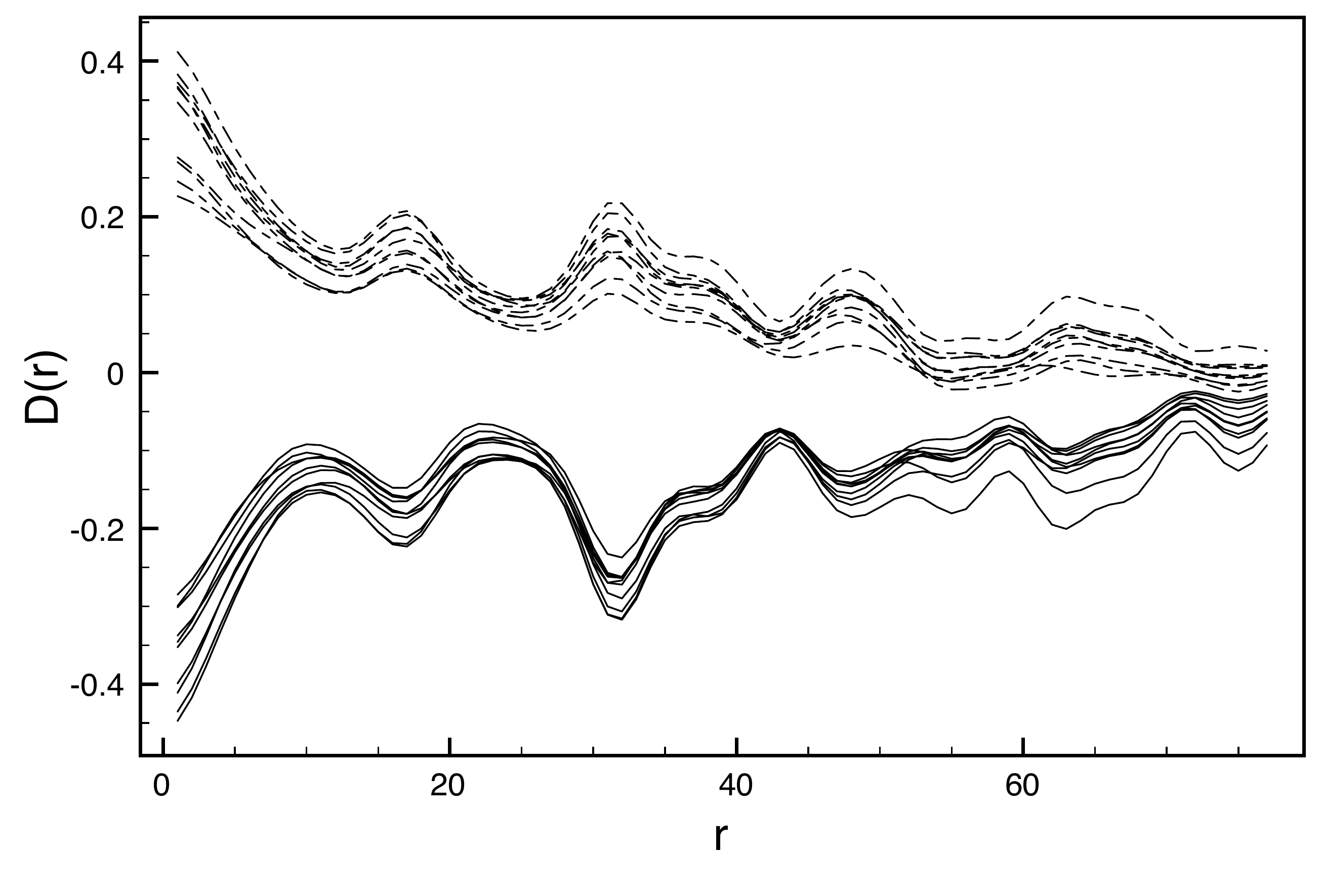}
\caption{Descriptors curves from each texture image. Solid curves correspond to sample 1. Dashed ones correspond to sample 2.}
\label{fig:result} 
\end{figure}

Figure \ref{fig:result} shows the descriptor curves for each image from TiO$_{2}$ samples. Each line aspect (solid and dashed) corresponds to a sample prepared in a different experimental condition.

It is noticeable that descriptors \textcolor{black}{curve} aspects \textcolor{black}{discriminate} strongly the materials. We do not observe any interlacement among curves from different conditions. The whole graphic reflects a high inter-class variability and a low intra-class variation in descriptors. Such result is in accordance to the pixel arrangement nature, measured through fractal descriptors.

Moreover, we notice the complex shape of each curve, reflecting the richness of information expressed in the descriptors. We observe a clear improve in the description compared to the classical fractal dimension.

\begin{figure}[!h]% Figuras lado a lado
\centering 
\includegraphics[width=0.45\textwidth]{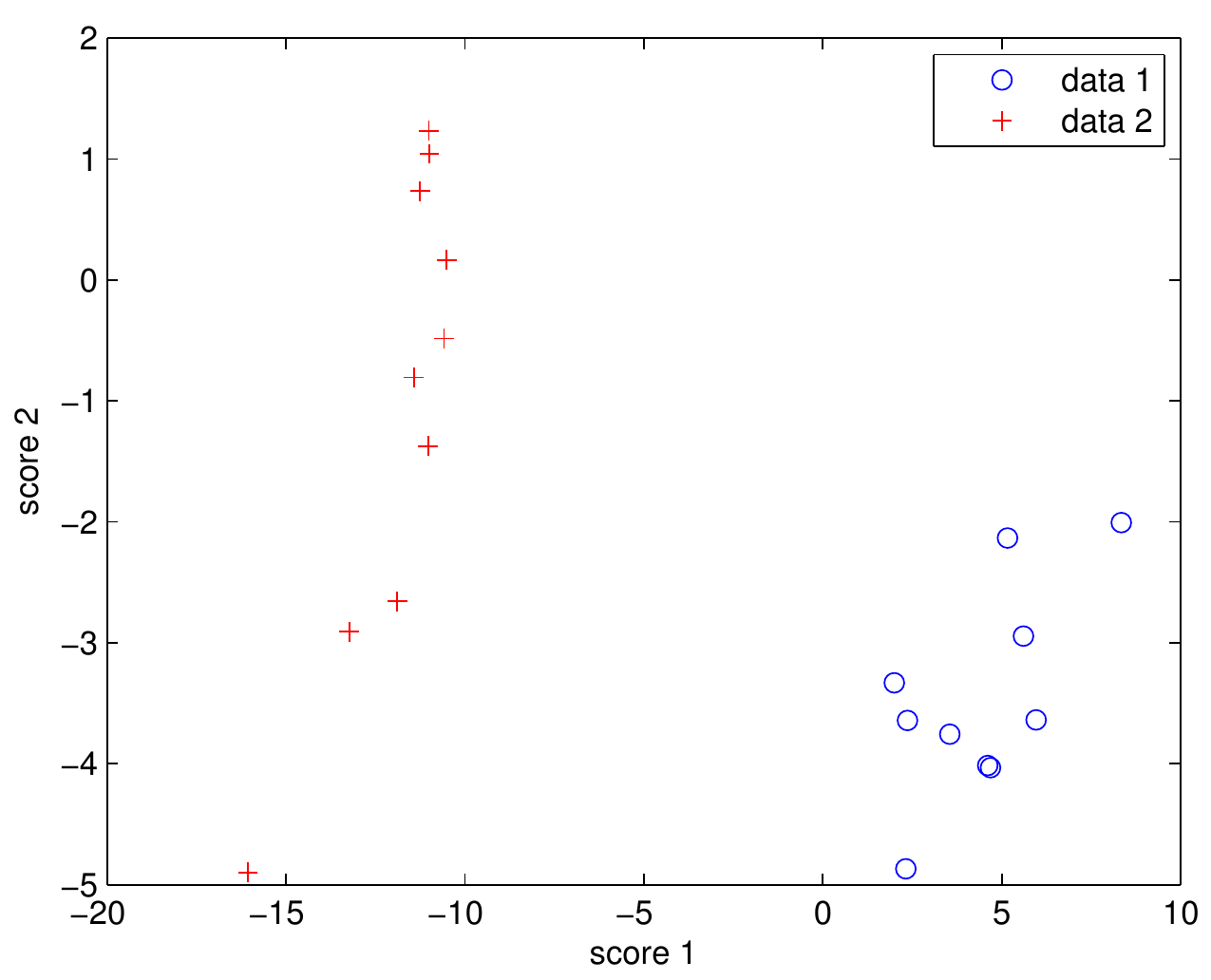}
\includegraphics[width=0.45\textwidth]{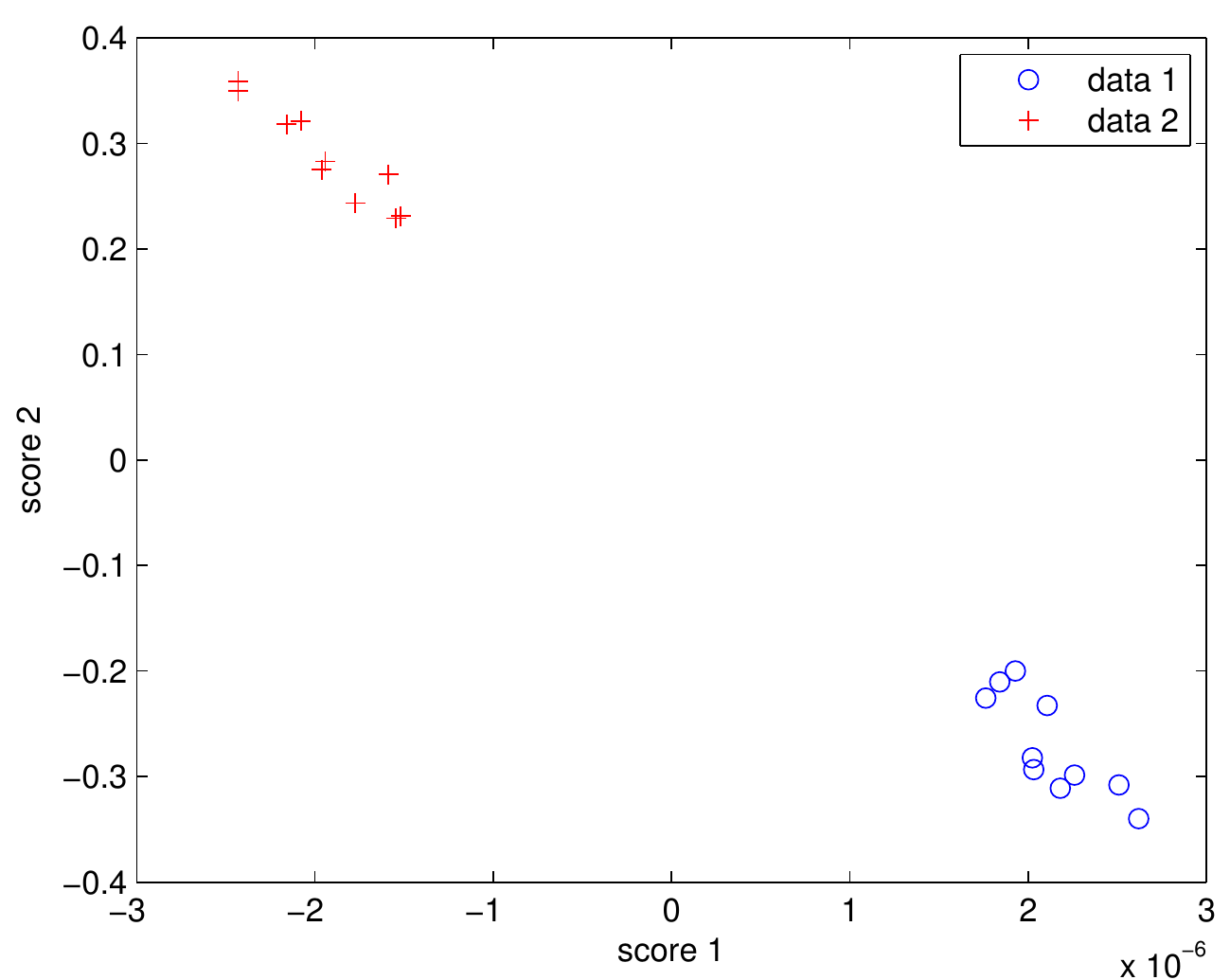}
\caption{Distribution of correlation scores. (a) Principal Component Analysis. (b) Canonical Correlation Analysis.}
\label{fig:corr} 
\end{figure}

Figure \ref{fig:corr} shows the distribution of two statistical important scores of fractal descriptors: from Principal Component Analysis (PCA) and \textcolor{black}{Canonical Correlation Analysis (CCA)}. %

While PCA plot measures the dispersion of whole dataset, CCA measures the correlation among data from a same material. Both graphics show
a cluster of points (corresponding to the 2 main scores) in each material conditions, comprobing again the discrimination power of fractal descriptors.

\section{Conclusions}

In this \textcolor{black}{work,} we proposed the application of fractal descriptors to the discrimination of materials under different experimental conditions. We obtained the descriptors by applying a multiscale approach to the fractal dimension \textcolor{black}{estimation} of texture images extracted from the material.

The curves of descriptors showed a high accuracy in the discrimination. This result demonstrated the power of the proposed technique in the modelling of materials. It also comprobed the validity of materials analysis through computational texture analysis.

\end{document}